\documentclass{iaus}
\usepackage{graphicx}

\title[JD 11.~~Obscuring Supersoft X-ray Sources in Stellar Winds] 
{Obscuring Supersoft X-ray Sources in Stellar Winds.}

\author[Mikkel T.B. Nielsen, Carsten Dominik, Gijs Nelemans]   
{Mikkel T.B. Nielsen$^1$, Carsten Dominik$^{2,1}$,
 \and Gijs Nelemans$^1$}

\affiliation{$^1$Department of Astrophysics, IMAPP, Radboud University Nijmegen, \\ PO Box 9010, NL-6500 GL Nijmegen, the Netherlands \\ email: {\tt m.nielsen@astro.ru.nl} \\[\affilskip]
$^2$Sterrenkundig Instituut "Anton Pannekoek", Science Park 904, \\ 1098 XH, Amsterdam,
The Netherlands}

\pubyear{2011}
\volume{281}  
\pagerange{?}
\jname{Binary Paths to Supernova Ia Explosions}
\editors{A.C. Editor, B.D. Editor \& C.E. Editor, eds.}
\begin{document}

\maketitle

\begin{abstract}
We investigate the possibility of obscuring supersoft X-ray sources in the winds of companion stars. We derive limits on the amount of circumstellar material needed to fully obscure a 'canonical' supersoft X-ray source in the Large Magellanic Cloud, as observed with the Chandra X-ray Observatory.
\end{abstract}

\firstsection 

\section{Introduction}

As shown in \cite{van.den.Heuvel.et.al.1992}, a massive white dwarf (WD) accreting material from a companion star at the steady-burning rate given in \cite{Nomoto.1982} will emit X-rays with a spectrum consistent with that observed for supersoft X-ray sources (SSS). If steady burning WDs do indeed look like SSS and the single degenerate (SD) progenitor scenario is the dominant contributor to the SN Ia rate, then we should expect to see a corresponding population of SSS large enough to produce the observed SN Ia rate. However, as shown by \cite{Gilfanov.Bogdan.2010} and \cite{Di.Stefano.2010}, the number of observed SSS in both the Milky Way and external galaxies appears to be at least 1-2 orders of magnitude too low to account for the number of observed SNe Ia, both in spiral galaxies and ellipticals. This dearth of SSS may mean that the missing SSS are simply not there, and hence that they are not the dominant contributors to the SN Ia rate. An alternative possibility is that they do in fact exist and produce a significant fraction of the total SN Ia rate, but are somehow hidden from view of our current observational capabilities during their supersoft phase, only to become visible when they explode as SNe Ia.

In the following we wish to explore the latter option. The object has been to make a simple analytical model of a massive, accreting WD with a giant companion star which is losing matter into the circumstellar region (in addition to the matter it transfers to the accretor) to determine which combination of parameters renders the WD undetectable as a SSS. It should be noted that we do not necessarily claim that systems with the required parameters exist, or that they exist in large enough numbers to make up for the deficit in SSS, as compared to SN Ia rate. All we aim to do is get a general handle on how much circumstellar matter is needed to render a canonical SSS unobservable to modern X-ray telescopes. 

\section{Model}
We consider a massive ($\sim 1$ M$_{\odot}$) WD steadily accreting and burning material from a close companion star. The companion can be a main sequence, red giant or asymptotic giant branch star. The accretion mechanism is intentionally left unspecified, but can be envisioned to be any mechanism capable of supplying the WD with the required amount of matter, such as Roche-Lobe overflow (RLOF), Bondi-Hoyle wind accretion, tidally enhanced wind accretion (see \cite{Chen.et.al.2011}), and wind-RLOF accretion (see \cite{Mohamed.Podsiadlowski.2007}).

The binary is assumed to be embedded in a spherical distribution of matter that has been lost from the binary. In our simulations we have assumed the mechanism behind this mass loss to be a radiation driven wind from the donor star. However, other ways may be envisioned in which the system can lose material into the surroundings, such as binary interactions or thermal pulses from the donor. We note that there is observational support for the existence of circumstellar material in early type Ia spectra, see for example Sternberg's talk (Sternberg \textit{et al} 2011, in prep.). The material is assumed to be distributed spherically symmetrically around the donor star, somewhat analoguous to the wind bubble described in the talk by Chiotellis, see also \cite{Chiotellis.et.al.2011}, except that our model system is considered stationary with respect to the ISM.

The code is one-dimensional and assumes a line of sight that minimizes the amount of material between the accretor and the observer. The density of material at the surface of the WD, as well as the obscuring column from there to the observer, is derived by integrating a simple $r^{-2}$ density profile, where $r$ is the distance from the surface of the companion star emitting the obscuring wind. 

We assume solar chemical abundances and use the parametrized model of \cite{Wilms.et.al.2000}. The contribution to the obscuration from hydrogen and helium in the photon energy range relevant to Chandra (i.e. above $\sim$ 100 eV) is completely negligible. Also, the abundances and cross sections of iron group elements are too small to play a role. Therefore, what matters are K-shell ionizations of intermediate mass elements, in particular C, N and O.

As a concrete application we consider what the SSS will look like when located in the Large Magellanic Cloud (LMC) and observed with the ACIS-I detector on the Chandra X-ray Observatory.

Given an orbital separation between the WD and the companion we want to calculate what mass loss rate is required to obscure the system from X-ray observations. By the mass loss rate we mean the rate of mass that is lost from the companion star without being accreted by the WD, i.e. $\dot{M}_{total} = \dot{M}_{accr} + \dot{M}_{lost}$. This material will be present as a circumstellar wind bubble, potentially absorbing the X-rays from the SSS.

\section{Results}
\begin{figure}
\begin{center}
 \includegraphics[width=3.4in]{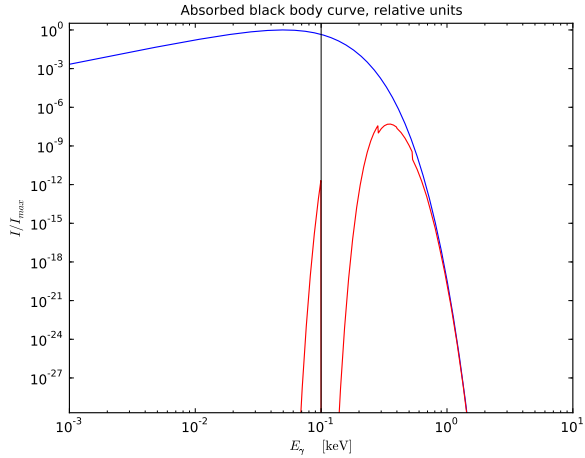} 
 \caption{Unabsorbed/naked (blue) and absorbed (red) black body curves of a canonical SSS ($L_{bol}=10^{38}$ erg/s, $kT_{peak}=50$ eV) in the LMC. The orbital separation is $a = 1$ AU and $\dot{M}=10^{-7}$ M$_{\odot}$/yr. The ordinate axis gives the relative number of photons arriving at the Earth.}
   \label{fig1}
\end{center}
\end{figure}

\begin{figure}
\begin{center}
 \includegraphics[width=3.4in]{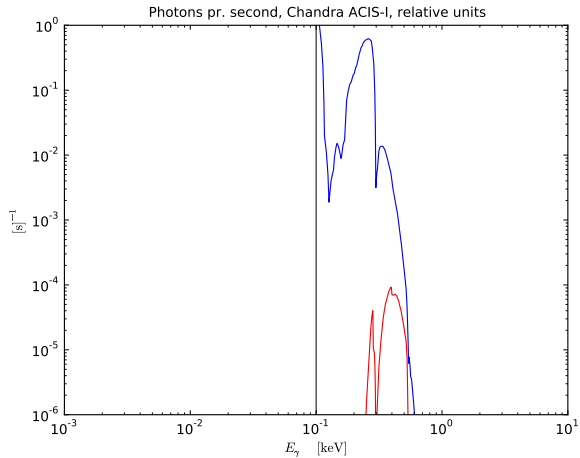}
 \caption{Relative number of photons registered on the ACIS-I detector of Chandra, as derived from the black body curves shown on figure \ref{fig1}. As in figure \ref{fig1} blue denotes the 'naked' SSS, while red is the absorbed spectrum. By our definition this source would be fully obscured.}
   \label{fig2}
\end{center}
\end{figure}

The black body curve of the unabsorbed and absorbed SSS is folded with the effective area function of ACIS-I detector on the Chandra satellite and the result is the relative number of photons received. Examples of this are shown in figures \ref{fig1}-\ref{fig2}. To have some measure of comparison we make a somewhat arbitrary work-in-progress definition that more than 2 orders of magnitude attenuation of the relative number of photons at an energy of 300 eV is considered 'full obscuration'.

The obscuration is dominated by the neutral material very close to the binary, while material further out contributes very little. This means that as long as the inner structure is sufficiently close to the spherical symmetry assumed in our model then our results should be applicable, since any additional structure further out plays a very negligible role in obscuring the source. Close to the X-ray source the hydrogen and helium will be ionized, and for these species only Thomson scattering contributes. This contribution is quite negligible. Intermediate mass elements can still be neutral close to the source, so K-shell ionizations of these species dominate the obscuration.

For $a \sim$ 1 AU the $\dot{M}_{lost}$ required for full obscuration is $\sim 10^{-7}$ M$_{\odot}$/yr, which is comparable to the steady-burning $\dot{M}_{accr}$ for a massive WD. The $\dot{M}_{lost}$ required for obscuration is approximately inversely proportional to the orbital separation (except for very small wind bubbles, where the outer extent of the wind bubble also plays a role), so as a rule of thumb, if the orbital separation is an order of magnitude larger, the mass loss rate also needs to be an order of magnitude larger to achieve the same amount of obscuration.

We find that for a given orbital separation there is a critical mass loss rate which determines the ionization structure around the source. For $\dot{M}_{lost} > \dot{M}_{crit}$ the hydrogen and helium is only ionized in a narrow region around the source. For $\dot{M}_{lost} < \dot{M}_{crit}$ the X-ray photons bleed out into the surrounding material, resulting in a large bubble of ionized hydrogen and helium. For a separation $a \sim$ 1 AU, the critical mass loss from the system is $\sim 10^{-6}$ M$_{\odot}$/yr. Thus, we can envision systems where the mass loss rate is large enough to obscure the X-rays but below $\dot{M}_{crit}$. The result will be systems that may be unrecognizable as SSS in X-rays, and have large ionized regions around them.

\section{Caveats}
Our wind model is probably the most important source of uncertainty in our model, and we have made a number of simplifying assumptions which may or may not have a significant effect on the applicability of our results.

Firstly, as mentioned earlier we have made a rather simplifying assumption concerning the line of sight between the observer and the WD, and this clearly underestimates the amount of obscuration that we expect to see from most sources.

Secondly, the assumption of constant wind speed from the surface of the companion star is probably not correct. In reality, the wind is accelerated by a variety of processes until it reaches its terminal velocity, and this is not expected to happen until well beyond the orbit of the binary. 
A larger wind speed means less obscuration, so our assumption probably underestimates the amount of obscuration.

Thirdly, we have assumed a spherically symmetric distribution of matter around the binary. This is in general not what is observed in symbiotic systems, where disk-like structures in the orbital plane are expected, possibly accompanied by polar outflows (see also talk by S. Mohamed). However, it is difficult to say whether our model is likely to over- or underestimate the amount of obscuration from the assumption of sphericality. 
Further studies of the geometry of accreting systems will provide a better understanding of the symmetries of the systems in question.

In general, then, two out of three of our more uncertain assumptions underestimates the amount of obscuration. In the case of the third assumption it is less clear whether it leads to an over- or underestimation of the obscuration, and more work is needed to clear up this point.

\section{Conclusions}
The implication of our work so far is that for LMC sources it appears possible to obscure a canonical SSS within the wind of a companion star, so that it is no longer observable as a SSS to Chandra's ACIS-I detector. In rough numbers, for orbital separations of $\sim 1$ AU mass loss rates of $\sim 10^{-7}$ M$_{\odot}$/yr are required to obscure a $kT_{peak}=50$ eV SSS in the LMC. In other words, if SD systems have wind mass loss rates comparable to the amount of matter accreted unto the WD component in the steady-burning case, then it is not unrealistic that such systems could become obscured to X-rays. However, unless the mass loss rates are very large ($\sim 10^{-6}$ M$_{\odot}$/yr for $a \sim 1$ AU) a persistent SSS will quickly create a large (several to hundreds of pc) ionized region around it, which will be observable in other wavelengths. Exactly how the X-rays will be recycled in the wind bubble is something we intend to explore in future work.

Our results will be reported in detail in an upcoming paper.



\begin{thebibliography}{}

\bibitem[Chen \etal\ (2011)]{Chen.et.al.2011}
{Chen, X., Han, Z., \& Tout, C.~A.} 2011,
\textit{Ap. Letters}, 735, L31

\bibitem[Chiotellis \etal\ (2011)]{Chiotellis.et.al.2011}
{Chiotellis, A., Schure, K.~M., \& Vink, J.} 2011,
\textit{ArXiv e-prints:} 1103.5487

\bibitem[Gilfanov \& Bogd{\'a}n (2010)]{Gilfanov.Bogdan.2010}
{Gilfanov, M. \& Bogd{\'a}n, {\'A}.} 2010,
\textit{Nature}, 463, 924

\bibitem[Greiner \etal\ (1991)]{Greiner.et.al.1991}
{Greiner, J., Hasinger, G., \& Kahabka, P.}, 1991
\textit{A\&A}, 246, L17

\bibitem[Hachisu \etal\ (2010)]{Hachisu.et.al.2010}
{Hachisu, I., Kato, M., \& Nomoto, K.} 2010,
\textit{Ap. Letters}, 724, L212

\bibitem[Helfand \& Grabelsky (1981)]{Helfand.Grabelsky.1981}
{Helfand, K.~S., Grabelsky, D.~A.} 1981,
\textit{ApJ}, 248, 925

\bibitem[van den Heuvel \etal\ (1992)]{van.den.Heuvel.et.al.1992}
{van den Heuvel, E.~P.~J., Bhattacharya, D., Nomoto, K., \& Rappaport, S.~A.} 1992,
\textit{A\&A}, 262, 97

\bibitem[Mohamed \& Podsiadlowski (2007)]{Mohamed.Podsiadlowski.2007}
{Mohamed, S. \& Podsiadlowski, P.} 2007,
\textit{ASP-CS}, 372, 397

\bibitem[Nomoto (1982)]{Nomoto.1982}
{Nomoto, K.} 1982,
\textit{ApJ}, 253, 798

\bibitem[Di Stefano (2010)]{Di.Stefano.2010}
{Di Stefano, R.} 2010,
\textit{ApJ}, 712, 728

\bibitem[Sternberg \etal\ (2011)]{Sternberg.et.al.2011}
{Sternberg, A., Gal-Yam, A., Simon, J.~D., Leonard, D.~C., Quimby, R.~M., Phillips, M.~M., Morrell, N., Thompson, I.~B., Ivans, I., Marshall, J.~L., Filippenko, A.~V., Marcy, G.~W., Bloom, J.~S., Patat, F., Foley, R.~J., Yong, D., Penprase, B.~E., Beeler, D.~J., Prieto, C.~A., and Stringfellow, G.~S.}, 2011
\textit{Science}, 333, 856

\bibitem[Tr{\"u}mper \etal\ (1991)]{Truemper.et.al.1991}
{Tr{\"u}mper, J., Hasinger, G., Aschenbach, B., Br{\"a}uninger, H., Briel, U.~G., Burkert, W., Fink, H., Pfeffermann, E., Pietsch, W., Predehl, P., Schmitt, J.~H.~M.~M., Voges, W., Zimmermann, U., \& Beuermann, K.}, 1991
\textit{Nature}, 349, 579

\bibitem[Wilms \etal\ (2000)]{Wilms.et.al.2000}
{Wilms, J. and Allen, A. \& McCray, R.}, 2000
\textit{ApJ} 542, 914




\end{thebibliography}
\end{document}